\documentclass[aps,pra,twocolumn,superscriptaddress]{revtex4-1} 

\usepackage{etex}
\usepackage{amsmath}
\usepackage{bm}
\usepackage{bbm}
\usepackage{listings}
\usepackage{graphicx}
\usepackage{graphics}
\usepackage{epsfig}
\usepackage{color}
\usepackage[dvipsnames]{xcolor}
\usepackage{multirow}
\usepackage[colorlinks]{hyperref}
\usepackage{fancyhdr}
\usepackage{calc}
\usepackage{natbib} 
\usepackage{bibentry}
\usepackage{bbm}
\usepackage{bbold}

\usepackage{mwe}
\usepackage[caption=false]{subfig}
\usepackage{float}
\DeclareMathOperator{\tr}{tr}
\newcommand{\dt}[1]{\frac{{\mathrm d} {#1}}{{\mathrm d}t}}
\def\br{\mathbf{r}}
\def\bra#1{\langle{#1}\rvert}
\def\ket#1{\lvert{#1}\rangle}

\newcommand{\nn}{\nonumber}
\newcommand{\mbf}[1]{\mathbf{#1}}

\newcommand{\smallfrac}[2]{\mbox{$\frac{#1}{#2}$}}

\newcommand{\expect}[1]{\big\langle #1 \big\rangle}

\newcommand{\AF}{A_{\rm Far}} 
\newcommand{\Ai}{A_{\rm in}} 

\usepackage{xspace}
\newcommand{\Poincare}{Poincar\'e\xspace}

\newcommand{\SWG}{square waveguide\xspace}

\begin{document}
\title{Enhanced cooperativity for quantum-nondemolition-measurement--induced spin squeezing of atoms coupled to a nanophotonic waveguide}
\author{Xiaodong Qi}
\email{qxd@unm.edu}
\affiliation{Center for Quantum Information and Control, University of New Mexico, Albuquerque, New Mexico 87131, USA}
\author{Yuan-Yu Jau}
\affiliation{Center for Quantum Information and Control, University of New Mexico, Albuquerque, New Mexico 87131, USA}
\affiliation{Sandia National Laboratories, Albuquerque, New Mexico 87185, USA}
\author{Ivan H. Deutsch}
\affiliation{Center for Quantum Information and Control, University of New Mexico, Albuquerque, New Mexico 87131, USA}
\date{\today}

\begin{abstract}
We study the enhancement of cooperativity in the atom-light interface near a nanophotonic waveguide for application to quantum nondemolition (QND) measurement of atomic spins.  Here the cooperativity per atom is determined by the ratio between the  measurement strength and the decoherence rate.  Counterintuitively, we find that by placing the atoms at an azimuthal position where the guided probe mode has the lowest intensity, we increase the cooperativity.  This arises because the QND measurement strength depends on the interference between the probe and scattered light guided into an orthogonal polarization mode, while the decoherence rate depends on the local intensity of the probe.  Thus, by proper choice of geometry, the ratio of good to bad scattering can be strongly enhanced for highly anisotropic modes. We apply this to study spin squeezing resulting from QND measurement of spin projection noise via the Faraday effect in two nanophotonic geometries, a cylindrical nanofiber and a square waveguide.  We find that, with about 2500 atoms and using realistic experimental parameters, $ \sim 6.3 $ and $ \sim 13 $ dB of squeezing can be achieved on the nanofiber and square waveguide, respectively. 
\end{abstract}

\maketitle

\section{Introduction}

Cooperativity is a measure of the entangling strength of the atom-light interface in quantum optics.  Originally introduced in cavity quantum electrodynamics (QED), the cooperativity per atom, $C_1$, can be expressed in terms of the ratio of the coherent coupling rate to the decoherence rates, $C_1 = g^2/(\Gamma_c \Gamma_A)$, where $g$ is the vacuum Rabi frequency,  $\Gamma_c$ is the cavity decay rate, and $\Gamma_A$ is the atomic spontaneous emission rate out of the cavity~\cite{Kimble1998}.  Alternatively, we can write $C_1 = (\sigma_0/A) \mathcal{F}$, where $\sigma_0$ is the resonant photon scattering cross section of the atom, $A$ is the cavity mode area, and $\mathcal{F}$ is the cavity finesse.  Expressed in this way, cooperativity is seen to arise due to scattering of photons preferentially into the cavity mode, compared to emission into free space, here enhanced by the finesses due to the Purcell effect. Strong coupling dynamics seen in pioneering experiments in atomic cavity QED~\cite{Raimond2001Manipulating, Miller2005} is now a mainstay in quantum information processing in systems ranging from quantum dots~\cite{Akimov2007, Akopian2006, Liu2010} to circuit QED~\cite{Wallraff2004Strong, Hofheinz2009Synthesizing}.  The $N_A$ atom cooperativity, $C_N = (N_A \sigma_0/A) \mathcal{F} =( OD) \mathcal{F}$, where $OD$ is the resonant optical depth.  In this configuration, the collective degrees of the atom can be manipulated by their common coupling to the cavity mode.

Cooperativity also characterizes the atom-light interface in the absence of a cavity.  In free space, an atom at the waist of a laser beam will scatter into the forward direction at a rate $\kappa \propto (\sigma_0/A) \gamma_s$, where $\gamma_s$ is the photon scattering rate into $4 \pi$ steradians~\cite{Baragiola2014}.  Here the single-atom cooperativity can be expressed to be proportional to the ratio of these rates, $C_1 \propto \kappa/\gamma_s \propto \sigma_0/A$.  The $N_A$-atom cooperativity, in a plane-wave approximation, ignoring effects of diffraction and cloud geometry~\cite{Baragiola2014}, $C_N \propto N_A \sigma_0/A = OD$.  To be self-consistent, here the beam area must be very large, so $C_1$ is very small, e.g., $C_1 \sim 10^{-6}$, but for a sufficiently large ensemble, the $OD$ can be large enough to lead to entanglement between the collective atomic degrees of freedom and the light.  In this situation, measurement of the light leads to back action on the ensemble and, for an appropriate quantum nondemolition (QND) interaction, results in squeezing of the collective spin~\cite{Kuzmich1998, Takahashi1999Quantum}. QND measurement-induced spin squeezing has been observed in free-space dipole-trapped ensembles~\cite{Appel2009Mesoscopic, Takano2009Spin, Sewell2012Magnetic} and in optical cavities~\cite{Schleier-Smith2010States, Cox2016Deterministic, Hosten2016}. The rate of decoherence is set by the rate of optical pumping,  $\gamma_{op} \propto \gamma_s$, and we can characterize the cooperativity per atom as $C_1 = \kappa/\gamma_{op}$.

In recent years nanophotonic waveguides have emerged as a new geometry that complements cavity QED, and can lead to strong cooperativity~\cite{Vetsch2010Optical, Chang2013,  Hung2013, Yu2014,  Douglas2015, Asenjo-Garcia2017Exponential, Qi2016}.  Notably, the effective beam area of a tightly guided mode can be much smaller than in free space and propagate for long distances without diffraction.  As such,  $\sigma_0/A$ can be orders of magnitude larger than in free space, e.g., $\sigma_0/A \sim 0.1$, and contribute collectively for a modest ensemble of a few thousand atoms trapped near the surface of the waveguide.  Moreover, in some cases the Purcell effect can further enhance forward scattering into the guided mode compared with scattering into free space.  Taken together, these features make  nanophotonic waveguides a promising platform for the quantum atom-light interface.  
 
In this paper we show that one can achieve an additional enhancement to the cooperativity in a nanophotonic geometry that is not possible in free space. In particular, we consider the QND measurement of the collective spin of an atomic ensemble via a Faraday interaction followed by polarization spectroscopy.  In this configuration the polarimeter effectively performs a homodyne measurement, where the probe is the ``local oscillator," which interferes with the light scattered into the orthogonally polarized guided mode~\cite{Baragiola2014}.   This signal thus depends on the spatial overlap of the two orthogonal polarization modes at the position of the atom.  In contrast, decoherence due to photon scattering into unguided $4 \pi$ steradians occurs at a rate $\gamma_s$ determined only by the intensity of the probe.   The net result is that the cooperativity per atom, $C_1 \propto \kappa/\gamma_s$, primarily depends on the strength of the {\em orthogonal polarization mode}, and this factor can be enhanced, especially for highly anisotropic guided modes.  Counterintuitively, we see that the strongest cooperativity arises when the atom is placed at the position of  minimum intensity of the azimuthally anisotropic probe mode where the intensity of the initially unoccupied orthogonal mode is maximum. 

We study the enhanced cooperativity for two nanophotonic geometries: a cylindrical nanofiber formed by tapering a standard optical fiber with cold atoms trapped in the evanescent wave, as recently employed in a variety of experimental studies~\cite{Vetsch2010Optical, Goban2012, Reitz2013, Lee2015, Goban2014, Reitz2014, Volz2014Nonlinear, Beguin2014, Mitsch2014,Kato2015Strong, Sayrin2015, Sayrin2015a, Mitsch2014a, Solano2017Dynamics, Beguin2017Observation}, and a nanofabricated suspended square waveguide, currently investigated at Sandia National Laboratories~\cite{Lee2017Characterizations}.  For each geometry we study the use of the Faraday effect and polarimetry to perform a QND measurement of the magnetic spins~\cite{Smith2003a}, and, thereby, induce squeezing of collective spins of cesium atoms.  A dispersive measurement of the number of atoms trapped near the surface of an optical nanofiber was first performed in~\cite{Dawkins2011}, and quantum spin projection noise was recently detected using a QND measurement with a two-color probe in~\cite{Beguin2014} and~\cite{Beguin2017Observation}.  Previously, we studied QND measurement-induced spin squeezing  mediated by a birefringence interaction~\cite{Qi2016}. We see here that, through the enhanced cooperativity, QND measurement via the Faraday effect can lead to substantial squeezing, greater than 10 dB in some geometries, for 2500 atoms.

The remainder of the paper is organized as follows.  In Sec. II, we lay out the theoretical description of the QND measurement and the relevant measurement strength.  In addition, we describe how decoherence is included in the model through a first-principles stochastic master equation, here for the case of alkali atoms, cesium in particular.  From this we see how cooperativity emerges as the key parameter that characterizes the squeezing.  We calculate in Sec. III the squeezing dynamics for the different nanophotonic waveguides, atomic preparations, and measurement protocols.  We conclude with a summary and outlook for future work in Sec. IV.

\section{QND measurement and cooperativity} \label{Sec::QNDandCooperativityTheory}
The theoretical framework describing the propagation of light guided in a nanofiber and interacting with trapped atoms in the dispersive regime was detailed in our previous work~\cite{Qi2016}.  We review the salient features here and include the generalization to the square waveguide.

\begin{figure}[htb]
\centering
  \includegraphics[width=.49\textwidth]{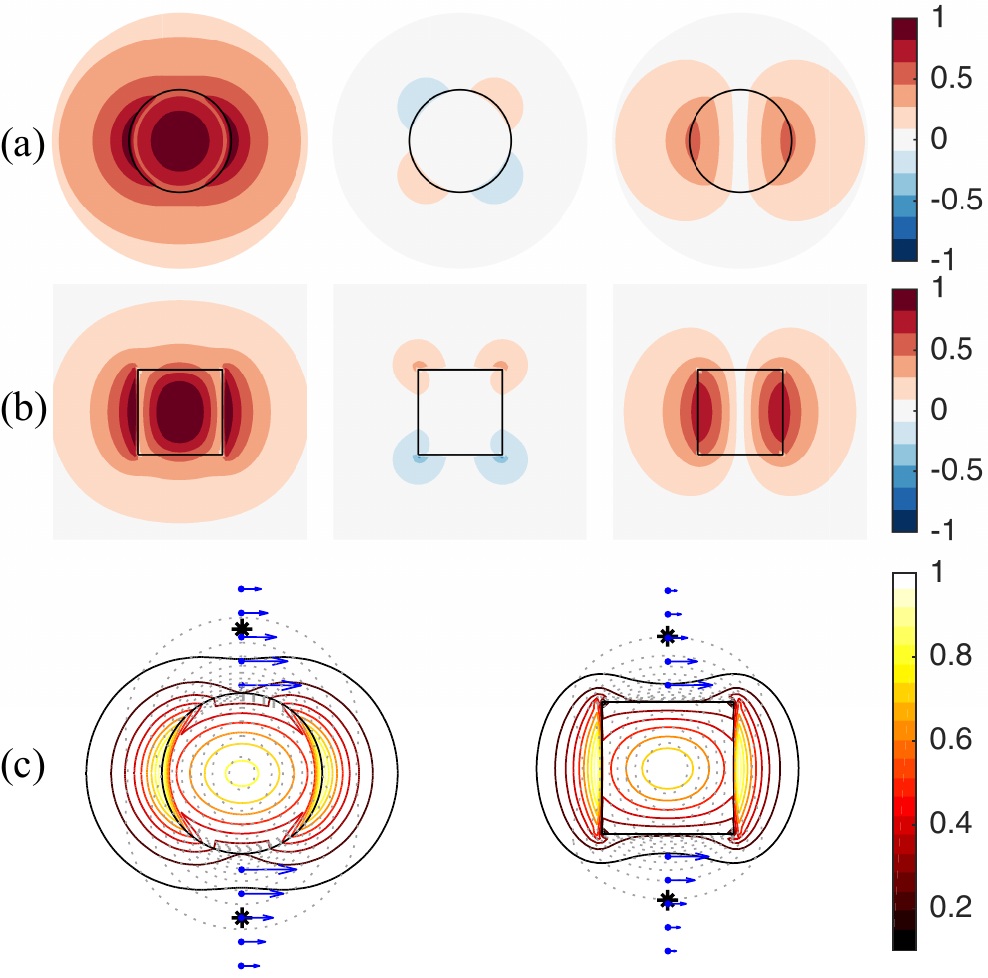}
  \caption{Fundamental guided modes of the nanophotonic waveguides. (a) Electric field components of the $H$-polarized HE$_{11}$ mode of a circular nanofiber. From left to right: $ \mathrm{Re}[u_x(\br\!_\perp)] $, $ \mathrm{Re}[u_y(\br\!_\perp)] $, and $ \mathrm{Im}[u_z(\br\!_\perp)] $ in the $ xy $ plane. (b) Same as (a) but for  the $H$-polarized quasi-TE$_{01}$ mode of a square waveguide. Black lines outline the waveguide boundary. The color scale is normalized to the maximum value of all field components for each waveguide mode. All other mode components not shown for both waveguide geometries vanish everywhere. (c) The normalized intensity distribution on the transverse plane for both geometries. Blue arrows indicate the local electric field's direction and amplitude (relative length) at positions along the vertical waveguide axis, which only have an $ x $-component of the mode. Stars indicate typical positions of trapped atoms, $r_\perp'/a=1.8  $ for the nanofiber~\cite{Vetsch2010Optical} and a similar scale for the \SWG,  $ r_\perp'/w=1.0 $, where $ a $ and $ w $ are the radius and width of the waveguides respectively). Dotted light gray lines show the corresponding $ V $-mode contour which is the $ H $ mode rotated by $ 90^\circ $ around the waveguide propagation axis. The atom's azimuthal position is chosen to be at a position with the $V$ mode being strongest. }\label{fig:nanofiberSWG_E_ints}
\end{figure}

For waveguides that are symmetric under a $\pi/2$ rotation around the $z$ (propagation) axis, there are two degenerate polarizations for each guided mode and for each propagation direction.  Assuming a nanophotonic waveguide that supports only the lowest order guided mode, and restricting our attention to modes propagating in the positive $z$-direction, we denote $\mbf{u}_H(\mbf{r}_\perp)$ and  $\mbf{u}_V(\mbf{r}_\perp)$ as the horizontally and vertically polarized modes that adiabatically connect to $x$ and $y$ linearly polarized modes, respectively, as the cross section of the waveguide becomes large compared to the optical wavelength.  Note that in typical nanophotonic geometries these guided modes also have a nonnegligible $z$ component.  For a cylindrically symmetric nanofiber, these are the well-studied HE$_{11}$ modes; for a \SWG, these are the quasi-TE$_{01}$ and quasi-TM$_{01}$ modes, shown in Fig.~\ref{fig:nanofiberSWG_E_ints}. One can solve for the guided modes of a cylindrical fiber analytically~\cite{Kien2004,Vetsch2010Opticala,Qi2016}. We use a vector finite difference method to numerically solve for the guided eigenmodes of the square waveguide~\cite{Fallahkhair2008} with core material of $ \rm{Si}_3\rm{N}_4 $ whose index of refraction is $ n=2 $~\cite{Lee2013}. 

The quasimonochromatic positive frequency component of the quantized field associated with these guided modes $(g)$ at frequency $\omega_0$ takes the form
\begin{align}\label{eq:Ebp}
\hat{\mathbf{E}}^{(+)}_g(\mbf{r}, t) &= \sqrt{ \frac{2 \pi \hbar \omega_0}{ v_g} } \left[\mathbf{u}_H(\mbf{r}\!_\perp)  \hat{a}_H(t) + \mathbf{u}_V(\mbf{r}\!_\perp) \hat{a}_V(t)\right]\nn\\
&\quad\quad\quad\quad\quad\cdot  e^{i (\beta_0 z- \omega_0 t)},
\end{align}
where $v_g$ is the group velocity, and $ \beta_0 $ is the propagation constant of the guided modes.  In the first Born approximation the dispersive interaction of the guided field with $N_A$ atoms trapped near the surface of the waveguide at positions $\{\mbf{r}'_\perp, z_n\}$, detuned far from resonance,  is defined by the scattering equation~\cite{Qi2016},
\begin{align}
\hat{\mathbf{E}}^{(+)}_{g,out}(\mbf{r}, t)&=\hat{\mathbf{E}}^{(+)}_{g,in}(\mbf{r}, t)\nn\\
&\quad+\sum_{n=1}^{N_A} \tensor{\mbf{G}}_{g} (\mbf{r}, \mbf{r}'_n;\omega_0) \cdot \hat{\tensor{\boldsymbol{\alpha}}}{}^{(n)} \cdot \hat{\mathbf{E}}^{(+)}_{g,in}(\mbf{r}'_n, t),
\end{align}
where $\hat{\tensor{\boldsymbol{\alpha}}}{}^{(n)}$ is the atomic polarizability operator of the $n$th atom, and 
\begin{equation}
		\tensor{\mathbf{G}}^{(+)}_g(\br,\br'_n; \omega_0) =  2\pi i \frac{\omega_0}{v_g } \sum_{p}\!\! \mathbf{u}_{p} (\br_\perp)\mathbf{u}^*_{p} 
(\br_{\perp}^\prime) e^{i \beta_0(z\!-\!z'_n)}  \label{Eq::GreensGuided}
\end{equation}
is the dyadic Green's function for a dipole to radiate into the forward-propagating guided mode.  In principle, the Green's function for an $N_A$-atom chain decomposes into collective sub- and superradiant normal modes~\cite{Asenjo-Garcia2017Atom,Asenjo-Garcia2017Exponential}, but in the far-detuning limit, these are all equally excited.  The result is equivalent to the symmetric mode of independently radiating dipoles.  The input-output relation for the mode operators then reads~\cite{Qi2016}
\begin{equation}
\hat{a}^{out}_p(t) = \hat{a}^{in}_p(t)  +i \sum_{p'} \hat{\phi}_{p,p'} \hat{a}^{in}_{p'}(t) ,
\end{equation}
where
\begin{equation}
\hat{\phi}_{p,p'} = 2\pi \frac{\omega_0}{v_g} \mbf{u}^*_p (\mbf{r}'_\perp) \cdot \sum_{n=1}^{N_A} \hat{\tensor{\boldsymbol{\alpha}}}{}^{(n)} \cdot \mbf{u}_{p'} (\mbf{r}'_\perp)
\end{equation}
is the phase operator associated with scattering polarization $p' \rightarrow p$ by a collective atomic operator.  When $p=p'$, this is a phase shift; for $p \neq p'$, this leads to a transformation of the polarization of the guided mode.

The Faraday effect arises from the irreducible rank 1 (vector) component of the polarizability tensor~\cite{Deutsch2010a}.  Given an atom with hyperfine spin $f$, this contribution is $\hat{\alpha}^{vec}_{ij} = i \alpha_1 \epsilon_{ijk} \hat{f}_k$, where $\alpha_1 = C^{(1)}_{f}\frac{\sigma_0}{4\pi k_0}\frac{\Gamma_A}{2\Delta} $ is the characteristic polarizability.  In alkali atoms $C^{(1)}_f=\mp\frac{1}{3f}$ for the D1- and D2-line transitions, respectively.  We take the detuning, $ \Delta $, large compared to the excited-state hyperfine splitting.  The resonant scattering cross section on a unit oscillator strength is $\sigma_0 = 6\pi/k_0^2$, where $k_0=\omega_0/c$.  The polarization transformation associated with scattering from $H$ to $V$ mode is determined by the operator
\begin{equation}
\hat{\phi}_{VH} = i 2\pi \frac{\omega_0}{v_g}\alpha_1 \left[ \mbf{u}^*_V (\mbf{r}'_\perp) \times  \mbf{u}_{H} (\mbf{r}'_\perp) \right] \cdot \hat{\mbf{F}},
\end{equation}
where $\hat{\mbf{F}}=\sum_n \hat{\mbf{f}}^{(n)}$ is the collective spin of the atomic ensemble.  Thus,
\begin{align}
\hat{a}^{out}_V(t) &= \hat{a}^{in}_V(t)  +i  \hat{\phi}_{V,H} \hat{a}^{in}_{H}(t)\nn\\
&= \hat{a}^{in}_V(t) \!-\! 2\pi\! \frac{\omega_0}{v_g}\alpha_1\! \left[ \mbf{u}^*_V (\!\mbf{r}'_\perp\!) \!\!\times\!\!  \mbf{u}_{H}(\!\mbf{r}'_\perp\!)\right]  \!\cdot\! \hat{\mbf{F}}\, \hat{a}^{in}_{H}(t),\label{eq:aoutain}
\end{align}
 and similarly for scattering from $V$ to $H$.

\begin{figure}[t]
\centering
  \includegraphics[width=.45\textwidth]{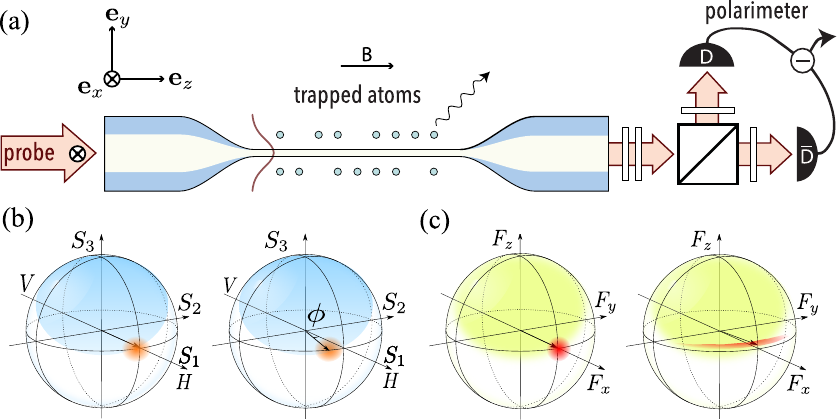}
  \caption{(a) Schematic polarization spectroscopy geometry for the QND measurement and spin-squeezing generation based on the Faraday effect.  Atoms trapped near the surface of the nanophotonic waveguide cause a Faraday rotation of the guided light, which is measured in a polarimeter that detects the $S_2$ component of the Stokes vector (intensity in the diagonal $D$ minus antidiagonal $\bar{D}$ modes).  (b) The evolution of the light's polarization state on the \Poincare sphere  (left to right).  The Stokes vector of the light is prepared along the $S_1$ direction, and the Faraday interaction causes a rotation around the $S_3$ axis.  Shot noise, shown as the uncertainty bubble, limits the resolution of the detection. (c) Evolution of the collective state before and after measurement (left to right).  The spin is prepared in a coherent state, with projection noise shown as the uncertainty bubble.  After the measurement the uncertainty in $F_z$ is squeezed, and the direction is correlated with the measurement outcome on the polarimeter. }\label{fig:spinsqueezingschematic}
\end{figure}

The polarization transformation can be expressed as a rotation of the Stokes vector of the light on the \Poincare sphere with operator components
\begin{subequations}\label{Eq::StokesComponents}
	\begin{align}
		\hat{S}_1(t) & = \smallfrac{1}{2}\big[ \hat{a}^\dag_H(t) \hat{a}_H(t)-\hat{a}^\dag_V(t) \hat{a}_V(t) \big], \\
	 	\hat{S}_2(t) & = \smallfrac{1}{2}\big[ \hat{a}^\dag_H(t) \hat{a}_V(t)+\hat{a}^\dag_V(t) \hat{a}_H(t) \big], \\
		\hat{S}_3(t) & = \smallfrac{1}{2i}\big[ \hat{a}^\dag_H(t) \hat{a}_V(t) -\hat{a}^\dag_V(t) \hat{a}_H(t) \big].
	\end{align}
\end{subequations}
By measuring the output Stokes vector in a polarimeter, we perform a QND measurement of a collective atomic operator to which it was entangled.  In a proper configuration, this leads to squeezing of a collective spin.  Launching $H$-polarized light corresponds to the initial Stokes vector along $S_1$, and Faraday rotation leads to an $S_2$ component, which is measured in a polarimeter [Fig.~\ref{fig:spinsqueezingschematic}(a)].  Taking the $H$-mode as a coherent state with amplitude $\beta_H$, the signal of the polarimeter measures $\hat{S}_2^{out} = (\beta_H \hat{a}_V^{\dag out} +\beta^*_H \hat{a}_V^{out})/2$.  Using this expression we see that the polarimeter acts as a homodyne detector, with the input $H$ mode acting as the local oscillator and the photons scattered into the $V$ mode as the signal.  Formally, the input-output relation follows from the scattering equation, Eq.~\eqref{eq:aoutain}, and reads
\begin{equation}
\hat{S}^{out}_2 \!=\! \hat{S}^{in}_2 \!+\!i \big( \hat{\phi}_{VH}\!-\! \hat{\phi}_{HV} \big) \hat{S}^{in}_1 \!=\!  \hat{S}^{in}_2 \!+\! \chi_3(\mbf{r}'_\perp) \hat{F}_z \hat{S}^{in}_1.
\end{equation}
The first term $\hat{S}^{in}_2$ represents the shot-noise, which fundamentally limits the resolution of the measurement and thus the spin squeezing that can be obtained in a given time interval.  The second term is the homodyne signal, where we have expressed the rotation angle around the 3 axis of the \Poincare sphere as
\begin{align}
\chi_3(\mbf{r}'_\perp) &= -\frac{4 \pi \omega_0}{v_g} \alpha_1 \left\vert \text{Re} \left[ \mbf{u}^*_V (\mbf{r}'_\perp) \times \mbf{u}_H (\mbf{r}'_\perp) \right] \right\vert\nn\\ &=-C^{(1)}_f  \frac{\sigma_0}{A_{Far}(\mbf{r}'_\perp)} \frac{\Gamma_A}{2 \Delta}.
\end{align}
We emphasize here the dependence of the rotation angle on the position of the atom in the transverse plane, $\mbf{r}'_\perp$, assumed equal for all atoms in the chain.  In particular, $\chi_3(\mbf{r}'_\perp)$ depends on the {\em overlap} of $\mbf{u}_H (\mbf{r}'_\perp)$ and $\mbf{u}_V (\mbf{r}'_\perp)$, indicating atomic scattering of photons from the $H$ to the $V$ modes associated with the Faraday interaction.  We have characterized this overlap by an effective area that defines the Faraday interaction at the position of the atom,
\begin{equation}\label{eq:AFar}
\AF(\mbf{r}'_\perp) = \frac{1}{n_g \left\vert \text{Re} \left[ \mbf{u}^*_V (\mbf{r}'_\perp) \times \mbf{u}_H (\mbf{r}'_\perp) \right]\right\vert},
\end{equation}
where $n_g = c/v_g$ is the group index.  A more tightly confined (smaller) area corresponds to a stronger interaction.

By monitoring the Faraday rotation, we can perform a continuous measurement on the collective spin projection $\hat{F}_z$.  The ``measurement strength," which characterizes the rate at which we gain information and thereby squeeze the spin, is given by
\begin{equation}\label{eq:kappa}
\kappa = \left\vert \chi_3(\mbf{r}'_\perp) \right\vert^2 \frac{P_{\rm in}}{\hbar \omega_0},
\end{equation}
where $P_{in}$ is the input power transported into the guided mode.  The measurement strength is the rate at which photons are scattered from the guided $H$ to the guided $V$ mode.  Decoherence arises due to diffuse scattering into unguided modes and the accompanied optical pumping of the spin.  In principle, the photon scattering rate into $4\pi$ steradians is modified over free space due to the Purcell effect, but we neglect this correction here.  In the case of the nanofiber, this is a small effect at typical distances at which the atom is trapped~\cite{LeKien2005,Kien2008}.  For the square waveguide, we examine this correction in future work.  

Decoherence is due to optical pumping of the spin between different magnetic sublevels.  Henceforth, we restrict ourselves to alkali atoms driven on the D1 or D2 line, at optical pumping rate 
\begin{equation}
\gamma_{op} =\frac{2}{9} \sigma(\Delta) \frac{I_{\rm in}(\mbf{r}'_\perp)}{\hbar \omega_0}.
\end{equation}
Here $\sigma(\Delta) = \sigma_0 \frac{\Gamma_A^2}{4 \Delta^2}$ is the photon scattering cross section at the detuning $\Delta$ for a unit oscillator strength transition, and the factor of $2/9$ reflects the branching ratio for absorbing a $\pi$-polarized laser photon followed by spontaneous emission of a photon, causing optical pumping to another spin state. $I_{\rm in}(\mbf{r}'_\perp) = n_g P_{\rm in}\vert \mbf{u}_H (\mbf{r}'_\perp)  \vert^2 \equiv P_{\rm in}/\Ai(\mbf{r}'_\perp) $ is the input intensity into the guided $H$ mode at the position of the atom,  where we have defined
\begin{equation}\label{eq:Ain}
\Ai(\mbf{r}'_\perp) =  \frac{1}{n_g \vert \mbf{u}_H (\mbf{r}'_\perp) \vert ^2}
\end{equation}
to be the effective area associated with the input mode.  We thus define the cooperativity per atom
\begin{equation}\label{eq:C1}
C_1 (\mbf{r}'_\perp)  = \frac{\kappa}{\gamma_{op}} =\frac{\sigma_0}{2f^2} \frac{  \Ai(\mbf{r}'_\perp) }{[\AF(\mbf{r}'_\perp)]^2}.
\end{equation}
This is our central result.  Roughly, $1/[\AF(\mbf{r}'_\perp)]^2 \sim \vert \mbf{u}_V (\mbf{r}'_\perp) \vert ^2 \vert \mbf{u}_H (\mbf{r}'_\perp) \vert ^2$, thus $ C_1(\mbf{r}'_\perp) \sim \sigma_0 \vert \mbf{u}_V (\mbf{r}'_\perp) \vert ^2$. In the context of homodyne measurement, the signal to be measured is proportional to the overlap between the $ H $ and the $ V $ modes, while the decoherence rate depends on the intensity of the local oscillator or $ H $ mode. How large the initially {\em unoccupied} $ V $ mode is at the atoms' position determines the signal-to-noise ratio for a QND measurement.  We thus enhance the cooperativity by choosing the position of the atoms so that the {\em orthogonal}, unoccupied mode is large, while the intensity of the local input mode that causes decoherence is small.

We contrast this with squeezing arising from a birefringence interaction, as we studied in [24]. Linear birefringence corresponds to a relative phase between the ordinary and extraordinary linear polarizations, which can arise due to both the geometry of the anisotropic modes relative to the placement of the atoms, and the atoms' tensor polarizability. Here, the coupling is not optimal at the position of minimum intensity; it is maximum at the angle 45$^\circ$ between the $H$ and the $V$ modes. As such, one will not see as strong of an enhancement of the cooperativity as we find in our protocol employing the Faraday effect. 

\begin{figure}[htb]
\centering
\includegraphics[width=\linewidth]{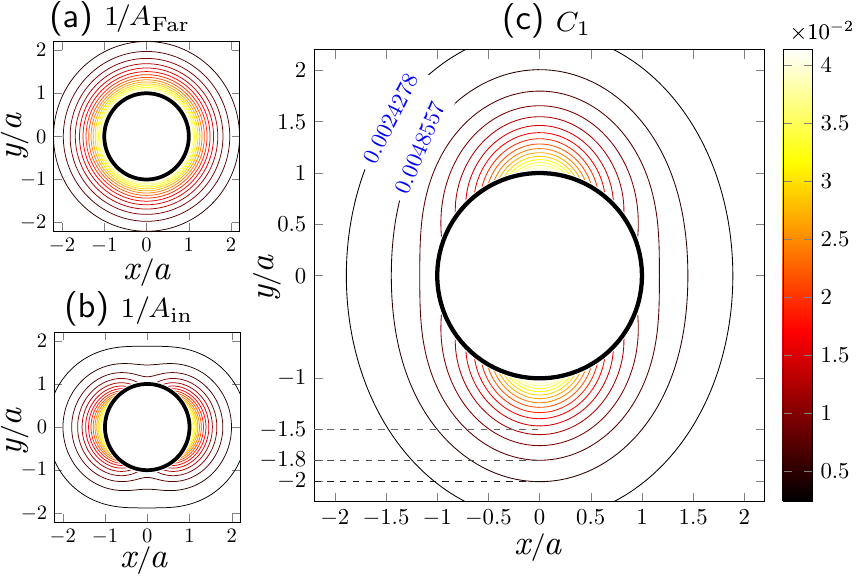}
\caption{Contour plots of the effective mode areas and the cooperativity per atom near an optical nanofiber. Contour plots of (a) reciprocal effective Faraday interaction mode area, Eq.~\eqref{eq:AFar},  and (b) the reciprocal input mode area in the $ xy $ plane, Eq.~\eqref{eq:Ain}. An $ x $-polarized incident mode is assumed. (c) Contour plot of the cooperativity, Eq.~\eqref{eq:C1}  in the $ xy $-plane. The isovalue lines of $ C_1 $ increase by $ 0.002428 $ at each gradient step from the outside inwards. The $ x$ and $y $ coordinates are scaled in units of $ a $ for all three plots.}\label{fig:nanofiber_Aeffgeometry}
\end{figure}

\begin{figure}[htb]
\centering
\includegraphics[width=\linewidth]{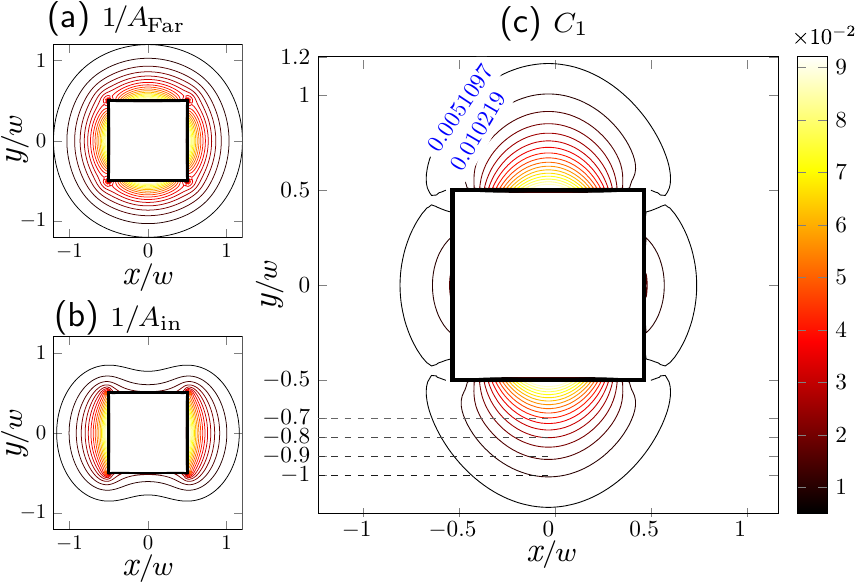}
\caption{Similar to Fig.~\ref{fig:nanofiber_Aeffgeometry}, but for the square waveguide. In (a), the contour lines outside of the waveguide are essentially concentric circles. There are distortions near the four corners of the square waveguide shown in the plot, mainly caused by numerical divergences. In (c), the isovalue lines of $ C_1 $ increase by $ 0.005109 $ at each gradient step from the outside inwards. The $ x$ and $y $ coordinates are scaled in units of $ w $ for all three plots.}\label{fig:square waveguide_Aeffgeometry}
\end{figure}

Figures~\ref{fig:nanofiber_Aeffgeometry} and~\ref{fig:square waveguide_Aeffgeometry} show plots of $1/\AF$, $1/\Ai$, and $C_1$ as a function of  the position of the atom in the transverse plane, $\mbf{r}'_\perp$, for the two nanophotonic geometries.  We see that $\AF$ is essentially cylindrically symmetric sufficiently far from the surface for both the nanofiber and the square waveguide geometries and thus the measurement strength is basically independent of the azimuthal position of the atoms.  In contrast, $\Ai$ is azimuthally anisotropic.  For input $x$ polarization, $1/\Ai$ is the smallest along the $y$ axis at a given radial distance, which corresponds to the lowest intensity of the input $H$ mode, and thus lowest optical pumping rate $\gamma_{op}$.  This angle corresponds to the position at which $\vert \mbf{u}_V (\mbf{r}'_\perp) \vert$ is largest and thus yields the largest enhancement of $C_1$.  Thus, counterintuitively, we enhance the cooperativity by placing the atom at the angle of minimum input intensity.   This enhancement is even greater for the square waveguide, which has more anisotropic guided modes compared to the cylindrical nanofiber.  For typical geometries, given a nanofiber with radius $a = 225$ nm and atoms trapped on the $y-$axis, a distance $200$ nm ($0.8a$) from the surface, the single-atom cooperativity is $C_1 =0.00728$ at the optimal trapping angle; for the square waveguide of width $w =300$ nm, with atoms trapped $150$ nm from the surface, $C_1 =0.0102$ at optimum.  Thus, with order $1000$ trapped atoms the $N_A$-atom cooperativity is of the order of $ 10 $, sufficient to generate substantial spin squeezing.

\section{Spin-squeezing dynamics}

Given an ensemble of $N_A$ atoms initially prepared in a spin-coherent state for the hyperfine spin $f$, polarized in the transverse plane, e.g., along the $x$ axis, a QND measurement of the collective spin  $F_z$ will squeeze the uncertainty of that component.  The metrologically relevant squeezing parameter is~\cite{Wineland1992},
\begin{align}\label{eq:xi2Faraday}
\xi^2 &\equiv  \frac{2 N_A f\expect{\Delta F_z ^2}}{\expect{\hat{F}_x}^2}.
\end{align}
Under the assumption that the state is symmetric with respect to the exchange of any two atoms, valid when we start in a spin-coherent state and all couplings are uniform over the ensemble, the collective expectation value can be decomposed into 
\begin{subequations}\label{eq:Ftof_squeezing}
\begin{align}
\expect{\!\Delta F_z^2} &= N_A \expect{\!\Delta f_z^2}\!+\!N_A(N_A\!-\!1)\left. \expect{\!\Delta\! f_z^{(i)}\Delta\! f_z^{(j)}}\right|_{i\!\neq\! j}\label{eq:DeltaFz2}\\
\expect{\hat{F}_x } & =N_A \expect{\hat{f}_x} \label{eq:expectFx}.
\end{align}
\end{subequations}
 The first term in Eqs.~\eqref{eq:DeltaFz2} and~\eqref{eq:expectFx} is the projection noise associated with the  $N_A$ identical spin-$f$  atoms, and  the second term in Eq.~\eqref{eq:DeltaFz2} is determined by two-body covariances, $ \left.\expect{\Delta f_z^{(i)}\Delta f_z^{(j)}}\right|_{i\neq j}=\expect{\Delta f_z^{(1)}\Delta f_z^{(2)}} = \expect{\hat{f}_z^{(1)}\hat{f}_z^{(2)}}-\expect{\hat{f}_z^{(1)}} \expect{\hat{f}_z^{(1)}} $.  Negative values of these two-body correlations correspond to the pairwise entanglement between atoms, leading to spin squeezing~\cite{Wang2003Spin}.  Note that when the detuning is sufficiently far off-resonance, all collective sub- and superradiant modes~\cite{Asenjo-Garcia2017Atom,Asenjo-Garcia2017Exponential} are equally (and thus symmetrically) excited.  In this paper, we work in the dispersive regime with a few thousand atoms and can safely ignore the atom-atom interaction caused by multiple scattering, and hence the collective atomic system satisfies the exchange symmetry. 

To study the spin-squeezing dynamics, we follow the method first developed by Norris~\cite{Norris2014}. We employ a first-principles stochastic master equation for the collective state of $N_A$ atoms, 
\begin{align}\label{eq:totaldrhodt}
\mathrm{d}\hat{\rho}= \left.\mathrm{d}\hat{\rho}\right|_{QND}+\left.\mathrm{d}\hat{\rho}\right|_{op}.
\end{align}
The first term on the right-hand side of Eq.~\eqref{eq:totaldrhodt} governs the spin dynamics arising from QND measurement~\cite{Jacobs2006,Baragiola2014},
\begin{align}
\left.\mathrm{d}\hat{\rho}\right|_{QND} &= \sqrt{\frac{\kappa}{4}}\mathcal{H}\left[\hat{\rho} \right]\mathrm{d}W + \frac{\kappa}{4}\mathcal{L}\left[ \hat{\rho}\right]\mathrm{d}t,
\end{align}
where  $\kappa$ is the measurement strength defined in Eq.~\eqref{eq:kappa}, and $\mathrm{d}W$ is a stochastic Wiener interval. The conditional dynamics are generated by superoperators that depend on the {\em collective} spin:
\begin{subequations}
\begin{align}
\mathcal{H}\left[ \hat{\rho}\right] &= \hat{F}_z \hat{\rho} + \hat{\rho}\hat{F}_z -2\expect{\hat{F}_z}\hat{\rho}, \\
\mathcal{L}\left[ \hat{\rho} \right] &= \hat{F}_z \hat{\rho}\hat{F}_z \!-\!\frac{1}{2}\!\left(\hat{\rho}\hat{F}_z^2\!+\!\hat{F}_z^2\hat{\rho} \right)\!=\!\frac{1}{2}\!\left[\hat{F}_z,\left[\hat{\rho},\hat{F}_z \right] \right].
\end{align}
\end{subequations}
The second term governs decoherence arising from optical pumping, which acts {\em locally} on each atom$,\mathrm{d}\hat{\rho}|_{op}=\sum_n^{N_A} \mathcal{D}^{(n)}\left[ \hat{\rho}\right] \mathrm{d}t$, where 
\begin{equation}
\mathcal{D}^{(n)}\!\left[ \hat{\rho}\right] = \!-\frac{i}{\hbar}\!\left(\!\hat{H}^{(n)}_{\rm eff}\hat{\rho} \!-\! \hat{\rho} \hat{H}^{(n)\dag}_{\rm eff}\right) \!+\! \gamma_{op}\!\! \sum_q\! \hat{W}^{(n)}_q\! \hat{\rho}\hat{W}^{(n)\dag}_q\!.
\label{op_superator}
\end{equation}
Here $\hat{H}^{(n)}_{\rm eff}$ is the effective non-Hermitian Hamiltonian describing the local light shift and absorption by the $n$th atom and $\hat{W}^{(n)}_q$ is the jump operator corresponding to optical pumping through absorption of a laser photon followed by spontaneous emission of a photon of polarization $q$~\cite{Deutsch2010a} (see Appendix~\ref{Sec::opticalpumpinginrotatingframe}).   

The rate of decoherence is characterized by the optical pumping rate, $\gamma_{op}$.  Note that the optical pumping superoperator, Eq.~\eqref{op_superator},  is not trace preserving when restricted to a given ground-state hyperfine manifold $f$.  Optical pumping that transfers atoms to the other hyperfine manifold in the ground-electronic state is thus treated as loss. Moreover, if the atoms are placed at the optimal position in the transverse plane, the local field of the guided mode is linearly polarized.  In that case the vector light shift vanishes, and for detunings large compared to the excited-state hyperfine splitting, the rank-2 tensor light shift is negligible. The light shift is thus dominated by the scalar component, which has no effect on the spin dynamics. In that case $\hat{H}_{\rm eff} = -i \frac{\hbar\gamma_{op}}{2} \hat{\mathbb{1}}$, representing an equal rate of absorption for all magnetic sublevels.

\begin{figure}[htb]
\centering
  \includegraphics[width=.35\textwidth]{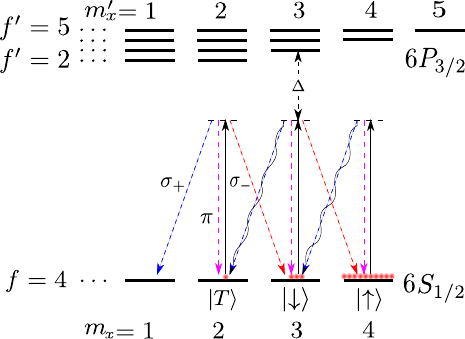}
  \caption{Schematic energy level diagram for cesium atoms probed on the D2 line, $6S_{1/2} \rightarrow 6P_{3/2}$.  Relevant dynamics are restricted to a truncated qutrit subspace of ground levels.  Atoms are prepared in the fiducial state $\ket{\uparrow}$, and driven by $x$-polarized ($\pi$) light.  The Faraday rotation corresponds to coherent scattering of $\pi \rightarrow \sigma$ in this basis, and measurement backaction leads to entanglement between pairs of atoms in $\ket{\uparrow}=\ket{f=4,m_x=4}$ and the coupled state $\ket{\downarrow}=\ket{f=4,m_x=3}$.  Optical pumping can cause spin flips $\ket{\uparrow} \rightarrow \ket{\downarrow}$ and $ \ket{\downarrow} \rightarrow  \ket{T}=\ket{f=4,m_x=2}$.  The latter process is included to account for transfer of coherences.  The detuning $\Delta$ is taken to be large compared to the excited-state hyperfine splitting.}
  \label{fig:spinsqueezinglevelstructure}
\end{figure}

Following the work of Norris~\cite{Norris2014}, the solution to the master equation is made possible by three approximations. First, we restrict the subspace of internal magnetic sublevels that participate in the dynamics.  The system is initialized in a spin-coherent state, with all atoms spin-polarized along the $x$ axis.  We denote this the ``fiducial state," $\ket{\uparrow} = \ket{f, m_x =f}$.   Through QND measurement, spin squeezing is induced by entanglement with the  ``coupled state,"  $\ket{\downarrow} = \ket{f, m_x=f-1}$.  Optical pumping is dominated by ``spin flips" $\ket{\uparrow}\rightarrow \ket{\downarrow}$ and ``loss" due to pumping to the other hyperfine level.  Finally, we include a third internal magnetic sublevel, $\ket{T} = \ket{f, m_x=f-2}$, to account for  ``transfer of coherences," which can occur in spontaneous emission~\cite{Norris2012Enhanced,Norris2014} (see Fig.~\ref{fig:spinsqueezinglevelstructure}).  Restricted to this qutrit basis, the internal hyperfine spin operators are
\begin{subequations}\label{eq:f_in_xbasis}
\begin{align}
\hat{f}_x &= f \hat{\sigma}_{\uparrow \uparrow} +(f-1) \hat{\sigma}_{\downarrow \downarrow} + (f-2)  \hat{\sigma}_{T T}, \\
\hat{f}_y &=-i \sqrt{\frac{f}{2}} \left(\hat{\sigma}_{\uparrow \downarrow} \!-\! \hat{\sigma}_{\downarrow \uparrow}\right) \!-i \sqrt{\frac{2f-1}{2}}  \left(\hat{\sigma}_{\downarrow T} \!-\! \hat{\sigma}_{T \downarrow }\right) \\
\hat{f}_z &= \sqrt{\frac{f}{2}} \left(\hat{\sigma}_{\uparrow \downarrow} + \hat{\sigma}_{\downarrow \uparrow}\right) + \sqrt{\frac{2f-1}{2}}  \left(\hat{\sigma}_{\downarrow T} + \hat{\sigma}_{T \downarrow }\right),
\end{align}
\end{subequations}
where we have defined the atomic population and coherence operators $\hat{\sigma}_{ba}=\ket{b}\bra{a}$.

Second, we assume that the collective state is symmetric under exchange of spins. This approximation is valid when all atoms see the same probe intensity and in the far-detuning regime when all sub- and superradiant modes are equally excited. With this, we can limit our attention to the symmetric subspace and define, for example, the symmetric two-body covariances by
\begin{align}
\expect{\!\Delta\sigma_{ba}^{(1)}\Delta\sigma_{dc}^{(2)}}_s \!\!\equiv\! \frac{1}{2}\!\left[\expect{\!\Delta\sigma_{ba}^{(1)}\Delta\sigma_{dc}^{(2)}}\!+\!\expect{\!\Delta\sigma_{ba}^{(2)}\Delta\sigma_{dc}^{(1)}} \right] ,
\end{align}
where the superscripts, (1) and (2), label arbitrarily two atoms in the ensemble. Due to the exchange symmetry, $ \expect{\Delta\sigma_{ba}^{(1)}\Delta\sigma_{dc}^{(2)}}_s=\expect{\Delta\sigma_{ba}^{(1)}\Delta\sigma_{dc}^{(2)}}=\expect{\Delta\sigma_{ba}^{(2)}\Delta\sigma_{dc}^{(1)}} $, which reduces the number of $ n $-body moments required to simulate the spin dynamics of the ensemble.

Third, we make the Gaussian approximation, valid for large atomic ensembles, so that the many-body state is fully characterized by one- and two-body correlations. Equivalently, the state is defined by the one- and two-body density operators, with matrix elements $\rho^{(1)}_{a, b} =\expect{\hat{\sigma}_{ba}}$, $\rho^{(1,2)}_{ac,bd}=\expect{\Delta \sigma_{ba}^{(1)}\Delta\sigma_{dc}^{(2)} }_s$ in the symmetric subspace.  We track the evolution of the correlation functions through a set of coupled differential equations~\cite{Norris2014}. Optical pumping, acting locally, couples only $n$-body correlations to themselves, e.g.,
\begin{align}
&\quad\left.d\expect{\!\Delta \sigma_{ba}^{(1)}\Delta\sigma_{dc}^{(2)} }_s\right|_{op} \nn\\
&= \expect{\mathcal{D}^\dagger[\Delta \sigma_{ba}^{(1)}]\Delta\sigma_{dc}^{(2)} }_sdt + \expect{\Delta \sigma_{ba}^{(1)} \mathcal{D}^\dagger[\Delta\sigma_{dc}^{(2)}] }_sdt .\label{eq:dsigmabadc_op}
\end{align}
QND measurement generates higher order correlations according to
\begin{equation}\label{eq:dsigmaba_QND}
\left.d\expect{\hat{\sigma}_{ba}}\right|_{QND} =\frac{\kappa}{4}\expect{\mathcal{L}^\dagger\left[\hat{\sigma}_{ba} \right]}dt + \sqrt{\frac{\kappa}{4}}\expect{\mathcal{H}^\dagger\left[\hat{\sigma}_{ba} \right]}dW .
\end{equation}
We can truncate this hierarchy in the Gaussian approximation, setting third-order cumulants to $ 0 $.  Thus, for example,
\begin{align}
&\quad\left.d\expect{\Delta \sigma_{ba}^{(1)} \Delta \sigma_{dc}^{(2)}}_s \right|_{QND}\nn\\
&= \left.d\expect{\hat{\sigma}_{ba}^{(1)} \hat{\sigma}_{dc}^{(2)}}_s \right|_{QND} - \left. \expect{\hat{\sigma}_{ba}} \right|_{QND} \left( \left.d\expect{\hat{\sigma}_{dc}} \right|_{QND}\right) \nonumber\\
&\quad -\! \left. \expect{\!\hat{\sigma}_{dc}\!}\! \right|_{QN\!D} \left(\! \left.d\expect{\!\hat{\sigma}_{ba}\!} \!\right|_{QN\!D}\!\right)
\!-\! \left.d\expect{\!\sigma_{ba}\!} \!\right|_{QN\!D}\left.d\expect{\sigma_{dc}} \right|_{QN\!D} \nonumber \\
&= -\kappa\expect{\Delta \sigma^{(1)}_{ba}  \Delta F_z }_s \expect{\Delta F_z \Delta \sigma_{dc}^{(2)} }_sdt,\label{eq:dsigmabadc_QND}
\end{align}
where we have employed the Ito calculus $dW^2 = dt$.
Note that when the third-order cumulants are set to $ 0 $,  the contribution of the $\mathcal{L}$ superoperator to dynamics of the two-body covariances vanishes,  $ \left.d\expect{\Delta \sigma_{ba}^{(1)} \Delta \sigma_{dc}^{(2)}}_s\right|_\mathcal{L} =\frac{\kappa}{4}\expect{\mathcal{L}^\dagger\left[\Delta\sigma_{ba}^{(1)}\Delta\sigma_{dc}^{(2)} \right]}_sdt=0 $.  This indicates the events of no-photon detection under the Gaussian-state approximation do not affect atom-atom correlations;  the measurement backaction and squeezing arise from the homodyne detection in the guided modes.

Using all of the approximations above, we can efficiently calculate the collective spin dynamics for the ensemble of qutrits (dimension  $d=3$) with $ d^2=9 $ equations for the one-body quantity, $ \expect{\hat{\sigma}_{ba}} $, and $ d^2(d^2+1)/2=45 $ equations for the two-body covariances, $ \expect{\Delta \sigma_{ba}^{(1)}\Delta\sigma_{dc}^{(2)} }_s $, in the symmetric subspace independent of the number of atoms.  With this formalism in hand, we can calculate the squeezing parameter, Eq.~\eqref{eq:xi2Faraday}, as a function of the time by finding time-dependent solutions for the one-body averages $\expect{\hat{f}_x}$ and  $\expect{\Delta f_z^2}$, and the two-body covariances $\expect{\Delta f_z^{(1)} \Delta f_z^{(2)}}$.  The detailed approach to calculating the collective spin dynamics is given in Appendix~\ref{Sec::opticalpumpinginrotatingframe}. 

\begin{figure}[htb]
\centering
\includegraphics[width=\linewidth]{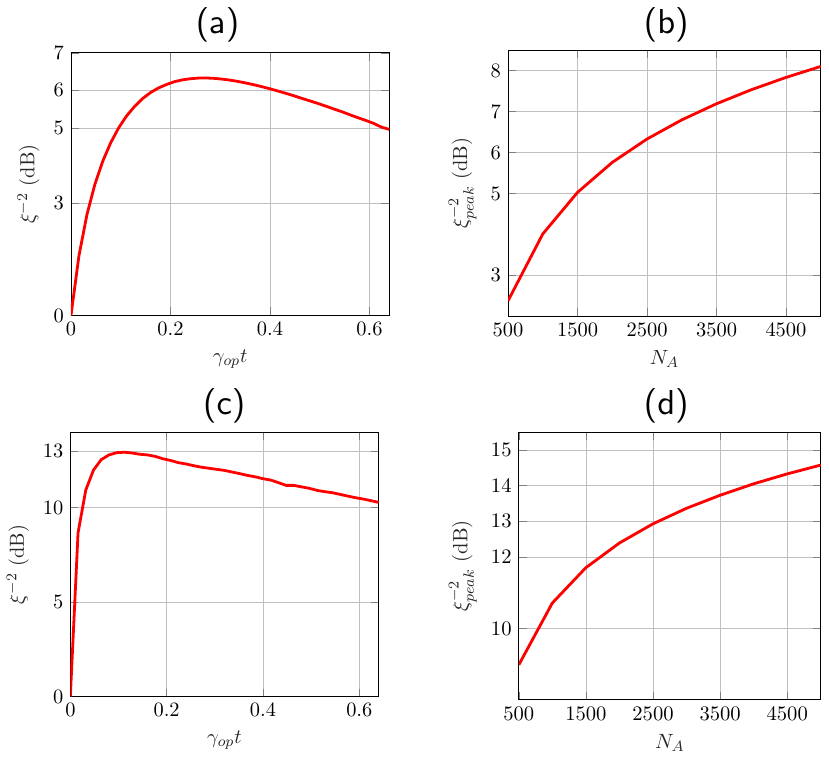}
\caption{ Reciprocal spin-squeezing parameter, Eq.~\eqref{eq:xi2Faraday}.  (a, c) Plots of $ \xi^{-2} $  as a function of time in units of the optical pumping rate $\gamma_{op}$, for the cylindrical nanofiber and square waveguide, respectively, for $N_A =2500$, with other parameters given in the text. These curves peak at the time determined by the detailed balance of reduced uncertainty due to QND measurement and decoherence due to optical pumping.  (b, d) Plots of the peak value $ \xi^{-2} $ as a function of $ N_A $ for the nanofiber and square waveguide, respectively. }\label{fig:xi_rpfix_NA_t}
\end{figure}

Using this formalism, we calculate the squeezing of an ensemble of cesium atoms, initially spin-polarized in the $6S_{1/2},\ket{f=4, m_x=4}$ state.  We choose the guided mode frequency near the D2 resonance, $6S_{1/2}\rightarrow 6P_{3/2}$, far detuned compared to the excited-state hyperfine splitting. In Fig.~\ref{fig:xi_rpfix_NA_t}, we plot the reciprocal of the spin squeezing parameter as a function of the time and its peak as a function of the atom number, $ N_A $, in both the optical nanofiber and the square waveguide cases. By placing atoms $ 200$  nm from the nanofiber surface ($ r'\!_\perp=1.8a $), our simulations for $ 2500 $ atoms yield $ 6.3 $ dB of squeezing. Using the square waveguide platform with the same number of atoms placed $150 $ nm from the surface, our calculation yields $12.9$ dB squeezing. As we have shown in Sec.~\ref{Sec::QNDandCooperativityTheory}, the square waveguide geometry enhances the anisotropic contrast of the two orthogonal guided modes and dramatically reduces the relative local intensity when the atoms are placed on the $ y $ axis. This results in a large cooperativity and higher peak spin squeezing, achieved in a shorter time and with a relatively slower decay compared to the nanofiber. 
In addition, in Figs.~\ref{fig:xi_rpfix_NA_t}(b) and~\ref{fig:xi_rpfix_NA_t}(d) we show how the peak squeezing scales with the number of trapped atoms when the atom positions are fixed as above. 

\begin{figure}[htb]
\centering
\includegraphics[width=\linewidth]{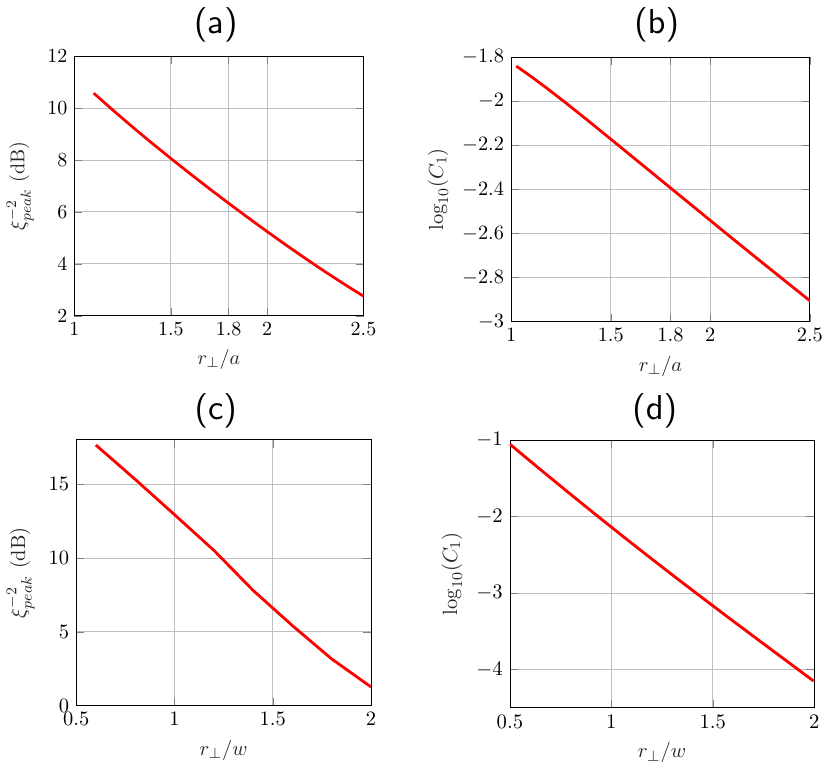}
\caption{(a, c) Peak squeezing parameter and (b, d) cooperativity at the optimal azimuthal trapping position as a function of the radial distance to the waveguide axis, for $N_A =2500$, with other parameters given in the text. Nanofiber case (a, b); square waveguide case (c, d). }\label{fig:peakxi_rp_NA}
\end{figure}

In the absence of any other noise, the cooperativity of atom-light coupling increases as the atoms are placed closer to the waveguide surface. Figures~\ref{fig:peakxi_rp_NA}(a) and~\ref{fig:peakxi_rp_NA}(c) show the peak squeezing  as a function of $ r\!_\perp $ for both the nanofiber and the square waveguide geometries with $2500$ atoms. With the same setting, we also plot out the cooperativity, $ C_1 $, on a logarithm scale in Figs.~\ref{fig:peakxi_rp_NA}(b) and~\ref{fig:peakxi_rp_NA}(d) as a function of the atom radial distance to the center of both waveguide geometries. We find that $C_1$ is proportional to $ e^{-\beta r\!_\perp} $ and the peak squeezing scales as $ \sqrt{OD} $, where $ \beta \approx 1.65/a $ for the nanofiber and $ \beta \approx 2.14\times 2/w $ for the \SWG with $2500$ atoms. The cooperativity of the \SWG increases more rapidly than that of the nanofiber as the atoms approach the waveguide surface.

\section{Summary and Outlook}
In this paper we have studied the cooperativity of the atom-photon interface for two nanophotonic geometries: a cylindrical nanofiber and a square waveguide.  Due to the anisotropic nature of the guided modes, one can strongly enhance the cooperativity by trapping atoms at positions that maximize the rate at which photons are forward-scattered  into the orthogonally polarized guided mode, while simultaneously minimizing the rate at which they are scattered into free space.  Counterintuitively, the optimal geometry is such that atoms at a certain distance from the surface are trapped at the azimuthal angle of the minimal intensity of the probe.  We applied this idea to study the generation of a spin squeezed state of an ensemble atoms induced by QND measurement, mediated by the Faraday interaction in the dispersive regime. 
With realistic parameters, our simulation shows more than $ 6 $ dB of squeezing for the cylindrical nanofiber or $ 12 $ dB for the square waveguide with $ 2500 $ atoms. The amount of spin squeezing we predict based on the Faraday effect is substantially larger than that for the birefringence-based spin-squeezing protocol studied in our earlier work~\cite{Qi2016}.  Although we have only considered a cylindrical nanofiber and a square waveguide, the ideas presented in this paper are applicable to other nanophotonic waveguide geometries which could show enhanced cooperativity.  In addition, our model of decoherence, simplified when the detuning is large compared to the excited-state hyperfine splitting, is almost certainly not the optimal operating condition.

Our simulations are based on a first-principles stochastic master equation that includes QND measurement backaction that generates entanglement and results in spin squeezing, as well as decoherence due to optical pumping by spontaneous photon scattering~\cite{Norris2014, Baragiola2014, Qi2016}. This simulation is made possible by a set of simplifying assumptions: (i) we restrict each atom to a qutrit, embedded in the hyperfine manifold of magnetic sublevels; (ii) the state is exchange symmetric with respect to any two atomic spins; and (iii) the many-body state is fully characterized by one- and two-body correlations (the Gaussian approximation).   With these, we solve for the metrologically relevant squeezing parameter as a function of time and see the tradeoffs between QND measurement and decoherence for various geometries and choices of parameters.  Our method is extendable to include higher order correlations, which become manifest at large squeezing, when the Holstein-Primakoff (Gaussian) approximation breaks down.  The computational framework we have developed here should allow us to study $n$-body moments with an acceptable computational load. 

In future work we intend to extend our analysis in a number of directions.  While we have focused here on the enhancement of the cooperativity in a nanophotonic waveguide-based QND measurement, we did not fully analyze the impact of the Purcell effect and the modification of spontaneous emission rates in our simulations.  We have also neglected here the motion of atoms in the optical lattice. In practice, however, these effects are important, the latter having been observed in the nanofiber  experiments~\cite{Solano2017Dynamics, Solano2017Alignment, Beguin2017Observation, Solano2017Optical}.  We include these in future studies, with an eye towards the development of new strategies for atomic cooling and state initialization in the nanophotonic platforms~\cite{Meng2017ground}. We expect that our proposed geometry, which places the atoms at positions of minimum intensity, could also help reduce the perturbation of the motion of trapped atoms due to the probe, which causes a disturbance in the signal~\cite{Solano2017Dynamics}.  

Finally, we have studied here the dispersive regime of a QND measurement, where the probe light is detuned far off-resonance, and multiple scattering  of photons among atoms is negligible. To extend our theory, it is necessary to include collective effects such as super- and subradiance~\cite{Asenjo-Garcia2017Exponential, Asenjo-Garcia2017Atom, Solano2017Super}, with applications including quantum memories~\cite{Sayrin2015, Gouraud2015Demonstration},  and the generation of matrix product states and cluster states~\cite{Economou2010, Lodahl2017Chiral, Schwartz2016Deterministic, Pichler2016Photonic, Pichler2017Photonic}.  


\section{Acknowledgments}
We thank Perry Rice for extensive discussions on sub- and superradiance and their effects on collective spin squeezing in the dispersive coupling regime and Ezad Shojaee for helpful discussions on the effects of decoherence on spin squeezing. We also thank Poul Jessen for insightful discussions regarding the enhancement of cooperativity. We acknowledge the UNM Center for Advanced Research Computing for computational resources used in this study. This work was supported by the NSF, under Grant No. PHY-1606989, and the Laboratory
Directed Research and Development program at Sandia National Laboratories.

Sandia National Laboratories is a multimission laboratory managed and operated by National Technology \& Engineering Solutions of Sandia, LLC, a wholly owned subsidiary of Honeywell International Inc. for the U.S. Department of Energy's National Nuclear Security Administration under contract DE-NA0003525.

%

\begin{appendix}
\section{Modeling collective spin dynamics}\label{Sec::opticalpumpinginrotatingframe}
In this Appendix we give further details of the equations of motion for spin squeezing as a function of the time, as discussed in the text, building on the work of Norris~\cite{Norris2014}.  In the symmetric subspace, and in the Gaussian approximation, we track a set of one- and two-body correlation functions that determine the metrologically relevant squeezing parameter, Eqs. (\ref{eq:xi2Faraday} and~\ref{eq:Ftof_squeezing}).   The evolution is determined by the dynamics induced by the QND measurement and decoherence due to optical pumping, Eqs. \eqref{eq:totaldrhodt}--\eqref{op_superator}.   For concreteness, we consider here cesium atoms,  initially prepared in a spin-coherent state, with all atoms in the stretched state polarized along the $x$ axis, $\ket{\uparrow} = \ket{6S_{1/2}, f=4, m_x=4}$.  The atoms are trapped on the $y$ axis at a distance $r'_\perp$ from the core axis of the waveguide and are probed with guided light in the $H$ mode, which has linear polarization in the $x$ direction at the position of the atoms.   We take the detuning large compared to the excited-state hyperfine splitting, e.g., 4 GHz red detuned from the D2 line,  $\ket{6S_{1/2}, f=4}\rightarrow \ \ket{6P_{3/2},f'=3}$.   Spontaneous emission from the probe may result in optically pumping of atoms to the other  hyperfine manifold, $f=3$; we treat this as a loss channel under the approximation that, over the time interval of interest, there is a negligible probability that these atoms will repump to $f=4$.  We also include a bias static magnetic field along the $z$ axis.  This does not affect the QND measurement of $F_z$, but ultimately, we must  calculate all dynamics in the rotating frame.

Spontaneous emission, optical pumping, and the resulting decoherence act locally on each atomic spin, governed by the master equation, Eq.~(\ref{op_superator}).  For light linearly polarized in the $x$ direction, and for detunings large compared to the excited-state hyperfine splitting, this takes the simplified form~\cite{Deutsch2010a}
\begin{align}
&\quad\left.\dt{\hat{\rho}^{(n)}}\right|_{op} =\mathcal{D}[\hat{\rho}^{(n)}]\nn\\
&=\!\!  -\frac{i}{\hbar}\![ \hat{H}\!\!_A, \hat{\rho}^{(\!n\!)}] \!\!-\! {\gamma_{op}}\hat{\rho}^{(\!n\!)} \!\!+\!\!\frac{\gamma_{op}}{4 f^2}\!\! \left(\!\hat{f}^{(\!n\!)}_+  \!\hat{\rho}^{(\!n\!)}\!  \hat{f}^{(\!n\!)}_- \!\!+\!\! \hat{f}^{(\!n\!)}_-\! \hat{\rho}^{(\!n\!)}\!  \hat{f}^{(\!n\!)}_+\!\right)\!\!,\label{eq:op_master}
\end{align}
where $\hat{f}^{(n)}_\pm = \hat{f}^{(n)}_z \pm i \hat{f}^{(n)}_y$ are the raising and lowering operators for projection of spin along the $x$ axis, and $\gamma_{op} = \frac{\Gamma_A^2}{18 \Delta^2}\;\sigma_0 \frac{I_{in}}{\hbar \omega_0}$ is the optical pumping for linear polarization in the far-detuned limit, given intensity $I_{in}$ at the position of the atom (we assume that all atoms are trapped the same distance for the waveguide and, thus, see the same intensity). The atomic Hamiltonian is $\hat{H}_A = \sum_n \hbar \Omega_0 \hat{f}^{(n)}_z + \hat{H}_{LS}$, the sum of the Zeeman interaction with a bias magnetic field, giving rise to Larmor precession at frequency $\Omega_0$, and the light shift, $\hat{H}_{LS}=\hbar \chi_3 \hat{F}_z \hat{S}_3$, due to the probe.  Even for far detuning, the residual tensor light shift cannot be neglected as it scales at $1/\Delta^2$, the same as $\gamma_{op}$ and $\kappa$.  In principle, a two-color probe can remove the tensor term which would otherwise lead to a degradation of the mean spin and, thus, a reduction in metrologically useful squeezing~\cite{Saffman2009,Montano2015Quantum}.  We neglect this effect here.  Finally, going to the rotating frame at the Larmor frequency, $\hat{f}^{(n)}_x \rightarrow \hat{f}^{(n)}_x \cos(\Omega_0 t) + \hat{f}^{(n)}_y \sin(\Omega_0 t)$,  $\hat{f}^{(n)}_y \rightarrow \hat{f}^{(n)}_y \cos(\Omega_0 t) - \hat{f}^{(n)}_x \sin(\Omega_0 t)$, and averaging the rapidly oscillating terms (RWA), the master equation takes the form
\begin{align}
&\quad\left.\dt{\hat{\rho}^{(n)}}\right|_{op}= \mathcal{D}[\hat{\rho}^{(n)}]\nn\\ 
&\Rightarrow  \!-{\gamma_{op}} \hat{\rho}^{(n)}\nn\\ 
&\quad +\!\frac{\gamma_{op}}{8 f^2}\! \left(\!\hat{f}^{(n)}_x  \!\hat{\rho}^{(n)}\! \hat{f}^{(n)}_x\!\!+\!\!\hat{f}^{(n)}_y  \!\hat{\rho}^{(n)}\!  \hat{f}^{(n)}_y \!\!+\!\!2 \hat{f}^{(n)}_z\!  \hat{\rho}^{(n)}\!  \hat{f}^{(n)}_z\!\right).\label{eq:op_master2}
\end{align}
With  atoms initially prepared in the  ``fiducial state," $\ket{\uparrow} = \ket{f=4,m_x=4}$, we include  in the dynamics the ``coupled state,"  $\ket{\downarrow} = \ket{f=4,m_x=3}$, and the ``transfer state," $\ket{T} = \ket{f=4,m_x=2}$.  Restricted to this qutrit substate, the spin projector operators $\{ \hat{f}_x,  \hat{f}_y, \hat{f}_z \}$ are given in Eqs.~\eqref{eq:f_in_xbasis}.

With all the components of the stochastic master equation defined in Eqs.~\eqref{eq:totaldrhodt}--\eqref{op_superator}, one can derive the equations of motion of one- and two-body moments straightforwardly using the symmetric Gaussian state approximation. Some explicit results have been given in Eqs.~\eqref{eq:dsigmabadc_op}--\eqref{eq:dsigmabadc_QND}. The equation of motion for the optical pumping dynamics of the one-body moment $ \expect{\hat{\sigma}_{ba}} $, is given by
\begin{align}
\left.d\expect{\!\hat{\sigma}_{ba}\!}\right|\!_{op}\!&=\! \expect{\!\mathcal{D}^\dagger\!\left[\hat{\sigma}_{ba} \right]\!}dt\!=\!\!\!\sum_{c,d}\!\!\tr(\mathcal{D}^\dagger\!\left[\hat{\sigma}_{ba}\right]\!\hat{\sigma}_{dc} )\expect{\!\hat{\sigma}_{dc}\!}dt.\label{eq:dsigmaba_op_expand}
\end{align}
The equations of two-body moments of the optical pumping can be derived similarly. Continuing from Eq.~\eqref{eq:dsigmabadc_op}, we have 
\begin{align}
&\quad\left.d\expect{\!\Delta \sigma_{ba}^{(\!1\!)}\Delta\sigma_{dc}^{(\!2\!)} }_s\right|_{op}\nn\\ 
&= \expect{\!\Delta\mathcal{D}^\dagger[ \sigma_{ba}^{(\!1\!)}]\Delta\sigma_{dc}^{(\!2\!)} }\!_sdt \!+\! \expect{\!\Delta \sigma_{ba}^{(\!1\!)} \Delta\mathcal{D}^\dagger[\sigma_{dc}^{(\!2\!)}] }\!_sdt\nn \\
&=\,\,\,\sum_{m,n}\! \tr\!\left(\mathcal{D}^\dagger[\hat{\sigma}_{ba} ]\hat{\sigma}_{mn}\! \right)\!\expect{\!\Delta\sigma_{mn}^{(\!1\!)}\Delta\sigma_{dc}^{(\!2\!)} }_s \nn\\
&\quad +\!\! \sum_{m,n}\!\tr\!\left(\mathcal{D}^\dagger[\hat{\sigma}_{dc}]\hat{\sigma}_{mn} \right)\!\expect{\!\Delta\hat{\sigma}_{ba}^{(\!1\!)}\Delta\sigma_{mn}^{(\!2\!)} }_s .\label{eq:dsigmabadc_op_expand}
\end{align} 
In deriving Eqs.~\eqref{eq:dsigmaba_QND} and~\eqref{eq:dsigmabadc_QND}, we have used the Gaussian state assumption to write three-body moments in connection to one- and two-body moments, $ \expect{\hat{A}\hat{B}\hat{C}}=\expect{\hat{A}\hat{B}}\expect{\hat{C}}+ \expect{\hat{A}\hat{C} }\expect{\hat{B}}+\expect{\hat{B}\hat{C} }\expect{\hat{A}}-2\expect{\hat{A}}\expect{\hat{B}}\expect{\hat{C}} $. 
If we keep only the coherence operators of the nearest coupling states, Eqs.~\eqref{eq:dsigmaba_op_expand} and~\eqref{eq:dsigmabadc_op_expand} recover the same results found by Norris~\cite{Norris2014}.

\allowdisplaybreaks
We apply similar techniques to derive the equations of motion due to QND measurement, Eqs.~\eqref{eq:dsigmaba_QND} and~\eqref{eq:dsigmabadc_QND}, to yield
\begin{align}\label{eq:dsigmaba_QND_expand}
\left.d\expect{\!\hat{\sigma}_{ba}\!}\right|_{QN\!D} &=\frac{\kappa}{4}\sum_{c,d}\tr\left(\mathcal{L}^\dagger\left[\hat{\sigma}_{ba} \right]\hat{\sigma}_{dc}\right) \expect{\hat{\sigma}_{dc}}dt \nn\\
&\quad +\!\! \sqrt{\frac{\kappa}{4}}\!\sum_{c,d}\!\tr\left(\mathcal{H}^\dagger\!\left[\hat{\sigma}_{ba} \right]\!\hat{\sigma}_{dc}\right)\! \expect{\!\hat{\sigma}_{dc}\!}dW .
\end{align}
\begin{align}
&\quad\left.d\expect{\!\Delta \sigma_{ba}^{(\!1\!)}\! \Delta \sigma_{dc}^{(\!2\!)}}\!_s \right|_{QN\!D} \nn\\
&=\!\! -\kappa\!\left\{\!\frac{1}{2}\!\left[ \sqrt{\frac{f}{2}}\!\left(\delta_{a\uparrow}\expect{\!\hat{\sigma}_{b\downarrow}\!}\!+\! \delta_{b\downarrow}\expect{\!\hat{\sigma}_{\uparrow a}\!}\!+\!\delta_{a\downarrow}\expect{\!\hat{\sigma}_{b\uparrow}\!} \!+\!\delta_{b\uparrow}\expect{\!\hat{\sigma}_{\downarrow a}\! } \right)\right.\right.\nn\\
&\quad\quad\,\quad\left. +\sqrt{\frac{2f\!-\!1}{2}}\!\left(\delta_{a\downarrow}\! \expect{\!\hat{\sigma}_{bT}\!} \!\!+\! \delta_{bT}\expect{\!\hat{\sigma}_{\downarrow a}\! } \!\!+\!\delta_{aT}\expect{\!\hat{\sigma}_{b\downarrow}\! } \!\!+\!\delta_{b\downarrow}\!\expect{\!\hat{\sigma}_{Ta}\! } \!\right)\!\right] \nn\\
&\quad\quad\quad -\!\sqrt{\!\frac{f}{2}}\expect{\!\hat{\sigma}_{ba}\! }\!\left(\!\expect{\!\hat{\sigma}_{\uparrow\downarrow}\! }\!\!+\!\expect{\!\hat{\sigma}_{\downarrow\uparrow}\! }\! \right)\! \!-\!\sqrt{\!\frac{2f\!\!-\!\!1}{2}} \expect{\!\hat{\sigma}_{ba}\!}\!\left(\!\expect{\!\hat{\sigma}_{\downarrow T}\!}\!+\!\expect{\!\hat{\sigma}_{T\downarrow}\!}\! \right)\nn\\ &\quad\quad\quad+\!(N_A\!-\!1)\left[\sqrt{\frac{f}{2}}\!\left(\expect{\!\Delta \sigma^{(1)}_{ba}  \Delta\sigma_{\uparrow\downarrow}^{(2)}}\!_s \!+\!\expect{\!\Delta\sigma_{ba}^{(\!1\!)}\Delta\sigma_{\downarrow\uparrow}^{(\!2\!)}}\!_s\right)\right.\nn\\
&\quad\quad\quad\quad\quad\quad\left.\left. + \sqrt{\frac{2f\!-\! 1}{2}}\!\left(\!\expect{\!\Delta \sigma^{(\!1\!)}_{ba}\Delta \sigma_{\downarrow T}^{(\!2\!)}}\!_s \!+\! \expect{\!\Delta\sigma_{ba}\Delta\sigma_{T\downarrow}^{(2)}}\!_s \!\right)\!\right] \! \right\}\nn\\
&\quad\!\cdot\!\left\{\! \frac{1}{2}\!\left[\! \sqrt{\!\frac{f}{2}}\!\left(\!\delta_{c\uparrow}\expect{\!\hat{\sigma}_{d\downarrow}\!}\!+\! \delta_{d\downarrow}\expect{\!\hat{\sigma}_{\uparrow c}\!}\!+\!\delta_{c\downarrow}\expect{\!\hat{\sigma}_{d\uparrow}\!} \!+\!\delta_{d\uparrow}\expect{\!\hat{\sigma}_{\downarrow c}\! }\! \right)\right.\right.\nn\\
&\quad\quad\quad\left. +\sqrt{\!\frac{2f\!-\!1}{2}}\!\left(\!\delta_{c\downarrow}\expect{\!\hat{\sigma}_{dT}\!} \!\!+\! \delta_{dT}\expect{\!\hat{\sigma}_{\downarrow c}\! }\!+\!\delta_{cT}\expect{\!\hat{\sigma}_{d\downarrow}\! } \!+\!\delta_{d\downarrow}\expect{\!\hat{\sigma}_{Tc}\! }\! \right)\!\right] \nn\\
&\quad\quad -\sqrt{\!\frac{f}{2}}\expect{\!\hat{\sigma}_{dc}\! }\!\left(\!\expect{\!\hat{\sigma}_{\uparrow\downarrow} \!}\!+\!\expect{\!\hat{\sigma}_{\downarrow\uparrow}\! }\! \right) \!-\!\sqrt{\!\frac{2f-1}{2}}\! \expect{\!\hat{\sigma}_{dc}\!}\!\left(\!\expect{\!\hat{\sigma}_{\downarrow T}\!}\!+\!\expect{\!\hat{\sigma}_{T\downarrow}\!}\! \right)\nn\\
&\quad\quad + \!(N_A\!-\!1)\!\left[\! \sqrt{\!\frac{f}{2}} \left(\expect{\Delta\sigma_{\uparrow\downarrow}^{(1)}\Delta\sigma_{dc}^{(2)}}_s \!+\!\expect{\Delta\sigma_{\downarrow\uparrow}^{(1)}\Delta \sigma_{dc}^{(2)}}_s\right)\right. \nn\\
&\quad\quad\left.\left. + \sqrt{\!\frac{2f\!\!-\!\! 1}{2}}\!\!\left(\!\expect{\!\Delta \sigma\!_{\downarrow T}^{(\!1\!)}\Delta\sigma\!_{dc}^{(\!2\!)}}\!_s \!\!+\!\expect{\!\Delta\sigma\!_{T\downarrow}^{(\! 1\! )} \Delta \sigma\!_{dc}^{(\!2\!)} }\!_s\!\right)\!\right]\!  \right\}\!dt,\label{eq:dsigmabadc_QND_expand}
\end{align}
By combining the optical pumping and QND measurement contribution to Eqs.~\eqref{eq:dsigmaba_op_expand}--\eqref{eq:dsigmabadc_QND_expand}, one can find a set of stochastic equations of a closed set of variables of 
\begin{equation}
\left\{\expect{\hat{\sigma}_{ba}},\expect{\Delta\sigma_{ba}\Delta\sigma_{dc} }_s\left|a,b,c,d\in \{\uparrow,\downarrow,T \}\right. \right\} 
\end{equation}
in the symmetric qutrit subspace. 
The matrix of equations is sparse and close to diagonal, which indicates that only nearest-neighbor coupling is possible in the $\left\{\expect{\hat{\sigma}_{ba}},\expect{\Delta\sigma_{ba}\Delta\sigma_{dc} }_s\right\}$ basis. 
In the symmetric qutrit subspace, we have $ 45 $ two-body moment variables and corresponding sparse second-order equations, which we solve numerically.

\end{appendix}
\end{document}